\documentclass{article}
\usepackage{graphicx} 
\usepackage{amsmath}
\usepackage{amsfonts}
\usepackage{amssymb}
\usepackage{float}
\usepackage{url}
\usepackage[toc,page]{appendix}
\usepackage{tabularx}
\usepackage{bbm}
\usepackage{xcolor}
\usepackage[margin=1in]{geometry}

\title{Prediction Markets Underperform Simple Baselines For Infectious Disease Forecasting}
\author{Carson Dudley and Reiden Magdaleno}
\date{\today}

\begin{document}

\maketitle

\begin{abstract}

Prediction markets (e.g., Polymarket, Kalshi) allow participants to bet on future events, producing real-time forecasts based on collective judgment. In domains such as elections and finance, markets have been effective at aggregating information, often rivaling or outperforming expert forecasters or polls. Whether this performance extends to infectious disease dynamics is unclear. Participants are self-selected and typically lack epidemiological expertise. However, markets can respond in real time to emerging news and unstructured signals in ways that standard forecasting pipelines cannot. Also, substantial financial stakes encourage participants to make an effort to be accurate. We evaluate Polymarket forecasts during 2025 and 2026 for two settings: weekly cumulative influenza hospitalizations in the US, which have an established expert-curated forecasting ensemble (CDC FluSight), and monthly measles cases, which do not. Across both settings, prediction markets fail to outperform standard benchmarks. For influenza, markets are competitive with low-performing individual FluSight models but are dominated by the FluSight ensemble: even when we combine market forecasts with the ensemble, the best combination puts zero weight on the markets. For measles, markets are outperformed by simple statistical baselines. We diagnose two sources of market inefficiency: placement of probability mass on impossible outcomes (e.g., decreasing values in cumulative forecasts) and low trading volume. These results suggest that current prediction markets are not reliable forecasters of infectious disease dynamics on their own or useful as complementary features for existing forecasting systems.
    
\end{abstract}

\section{Introduction}

Accurate forecasting of infectious disease dynamics is central to public health decision-making, informing resource allocation, policy interventions, and risk communication. Over the past decade, substantial progress has been made through a range of approaches, including mechanistic and statistical models \cite{desikan2022mechanistic, ihme2021modeling}, deep learning–based methods \cite{deepgleam, deepcovid}, and, more recently, foundation models for infectious disease forecasting that can forecast many diseases without modification \cite{mantis}. In practice, forecasts are often combined through ensembles, such as those used in the US Centers for Disease Control and Prevention (CDC) FluSight and COVID-19 Forecast Hubs \cite{flusight, forecasthub, ensembling, evaluationpnas, us_rsv_forecast_hub}, which consistently outperform individual models.

Despite their success, existing forecasting systems have limitations. Ensemble methods improve performance by averaging across models, but contributing models are often trained on similar epidemiological and surveillance data, potentially limiting diversity \cite{diversity}. Most forecasting pipelines also operate on fixed features and update at discrete intervals, which constrains their ability to incorporate new or unstructured information in rapidly evolving settings (e.g., news, social media posts, etc). These limitations have motivated interest in forecasting systems that can flexibly integrate heterogeneous, real-time signals.

Prediction markets represent one such system. In these markets (e.g., Polymarket, Kalshi), participants trade contracts tied to future outcomes, with prices reflecting collective expectations \cite{kalshi2026, polymarket2026}. Participants can incorporate any information they deem relevant, including local observations, expert judgment, and emerging news, and are financially incentivized to form accurate beliefs and to exploit discrepancies between their expectations and prevailing prices \cite{berg2003accuracy}. In domains such as elections and economic forecasting, prediction markets have demonstrated strong empirical performance and are often difficult to outperform \cite{snowberg2012prediction, berg2008prediction}. Whether this performance extends to infectious disease dynamics has received little empirical attention, despite growing interest in the markets themselves and the recent introduction of disease-related contracts.

The question is whether the mechanisms that make markets effective in other domains transfer to infectious disease forecasting. Elections and economic indicators involve discrete, well-defined outcomes that attract sustained participation from informed traders, and pricing often converges as resolution approaches \cite{berg2008prediction}. Disease forecasts, in contrast, involve continuous quantitative targets, resolve on shorter timescales, and attract participants whose access to epidemiological information is limited relative to the research groups contributing to expert hubs. Whether markets' ability to aggregate real-time information is enough to overcome these disadvantages is an open question.

In this work, we evaluate Polymarket forecasts during 2025 and 2026 for two infectious disease settings: weekly US influenza hospitalizations, which are the target of an established expert-curated ensemble (CDC FluSight), and monthly measles cases, where there is not a comparable ensemble. This design allows us to assess market performance both against a strong expert benchmark and in a setting where no such benchmark exists and markets could in principle provide useful forecasts.

Across both settings, we find that prediction markets fail to outperform standard benchmarks. For influenza, markets are competitive with lower-performing individual FluSight models but are dominated by the FluSight ensemble. Markets carry no information beyond what the ensemble already captures. When we combine them, the optimal mix puts zero weight on the markets. For measles, markets are outperformed by simple statistical baselines (e.g., ARIMA). We diagnose two sources of market inefficiency. Markets systematically place probability mass on impossible outcomes (e.g., for cumulative forecasts, on values below the most recently observed count) and this fraction, while declining over the flu season, never reaches zero. Trading volume is also orders of magnitude lower than in election markets where prediction markets have historically succeeded. These results clarify a question that has received limited empirical treatment in the infectious disease forecasting literature and argue against incorporating prediction market signals as complementary features in forecasting systems. We conclude by discussing why disease forecasting may differ from the domains where markets have succeeded, the ethical considerations raised by public health engagement with betting platforms, and the conditions under which this empirical question may be worth revisiting.

\section{Methods}

\paragraph{Prediction market forecasts} We evaluate Polymarket prediction markets for two infectious disease settings: weekly cumulative U.S. influenza hospitalization rates during the 2025--2026 season, and monthly cumulative U.S. measles case counts during 2025-2026. Influenza markets resolve to the cumulative influenza-associated hospitalization rate per 100,000 population as reported by the CDC FluSurv-NET system \cite{flusurv}, with each market corresponding to a specific epidemiological week. Measles markets resolve to the cumulative number of reported U.S. measles cases as of a specified end-of-month date, as published by the CDC National Notifiable Diseases Surveillance System \cite{cdc2026measles}.

Market structures differ between the two diseases. Influenza markets are structured as mutually exclusive range bins over hospitalization rate intervals, with participants trading contracts tied to each outcome. Measles markets are structured as a collection of threshold-based contracts of the form ``at least $N$ cases by [date]'' for a series of increasing thresholds for $N$. Each contract pays out if the cumulative case count meets or exceeds $N$ at resolution. We parse thresholds directly from contract labels and convert the resulting set of threshold probabilities into a discrete predictive distribution over case count bins by taking differences between adjacent threshold probabilities.

For both settings, let $p_i$ denote the normalized market-implied probability for bin $i$, obtained from contract prices. For each event, this defines a discrete predictive distribution over the target quantity. The realized outcome is mapped to the corresponding bin index based on the reported CDC value. If the realized value lies exactly on a bin boundary, the higher bin is selected (the same rules the prediction markets use).

To reconstruct market forecasts as they existed at each historical time point, we retrieve full price histories for every contract from the Polymarket API at 60-minute fidelity, spanning the full lifetime of each market. For measles, we reconstruct real-time CDC ground truth using archived snapshots of the CDC weekly measles surveillance, which allows us to match market prices at time $t$ with the surveillance data available at time $t$ rather than with final revised counts. This prevents look-ahead bias in the evaluation.

\paragraph{Evaluation metrics} We evaluate probabilistic forecasts using three proper scoring rules: the multiclass Brier score, the log score, and the continuous ranked probability score (CRPS). For each event, let $p$ denote the vector of predicted probabilities over bins and let $y$ denote the realized bin index. The Brier score,
\[
\text{Brier}(p, y) = \sum_i (p_i - \mathbbm{1}\{i = y\})^2,
\]
measures squared error between the predicted distribution and the realized outcome. The log score,
\[
\text{LogScore}(p, y) = -\log p_y,
\]
penalizes probability assigned away from the realized bin and is sensitive to sharp mispredictions (i.e., confidently wrong).

These metrics were chosen instead of the more standard weighted interval score (WIS) and mean absolute error (MAE) that are commonly used to evaluate infectious disease forecasts \cite{flusight, wis}, because prediction markets produce forecasts over a fixed set of discrete bins rather than quantile forecasts or continuous predictive distributions. Bin-based metrics operate directly on the native output of the markets, whereas WIS requires quantile-based forecasts that would need to be reconstructed from the bin probabilities via interpolation, introducing assumptions that are not part of the market's actual forecast.

This evaluation format advantages prediction markets relative to the baseline models. FluSight models and the statistical baselines we consider are designed to produce continuous or quantile forecasts, and converting them to the discrete bin structure means they are being evaluated on metrics they were not explicitly optimized for. Our findings are therefore, if anything, slightly biased toward the prediction markets.

\paragraph{Baseline forecasts} For influenza, we compare prediction markets to forecasts from the CDC FluSight Forecast Hub \cite{flusight}, including both individual models and the FluSight ensemble. We use one-week-ahead incident hospitalization forecasts from participating models and convert them to cumulative hospitalization rate predictions to match the prediction market target. Quantile forecasts are mapped to the prediction market bin structure by constructing a piecewise cumulative distribution function from the submitted quantiles and computing the implied probability mass within each bin. For measles, no comparable forecasting hub exists. We therefore compare prediction markets to an auto-ARIMA baseline fit to weekly cumulative case counts, with model order selected by AIC \cite{aic}. Predictive distributions are obtained from the fitted model's forecast distribution and discretized to the market bin structure using the same method as the FluSight forecasts.

\paragraph{Forecast alignment} Prediction markets update continuously, whereas FluSight forecasts are submitted weekly and baseline statistical models are refit each month. We use the full trajectory of market prices over the lifetime of each market, averaging scores across all available market snapshots to produce a single performance measure per event. This approach uses strictly more information than is available to the baseline models at their submission times, as market prices at later points in the market incorporate surveillance data and other signals that were not available when baseline forecasts were generated. Our evaluation therefore further advantages prediction markets, and any performance gap in favor of the baselines cannot be attributed to any asymmetry that disadvantages the markets.

\paragraph{Study period and sample} Prediction markets for influenza hospitalizations were introduced in January 2026. We analyze forecasts from January through May 2026, covering 16 markets at the time of analysis. For measles, we analyze markets from May 2025 through April 2026. All available markets within this period are included without filtering.

\section{Results}

\subsection{Influenza markets are outperformed by the FluSight ensemble and most individual models}

\begin{figure}[H]
\centering
\includegraphics[width=\textwidth]{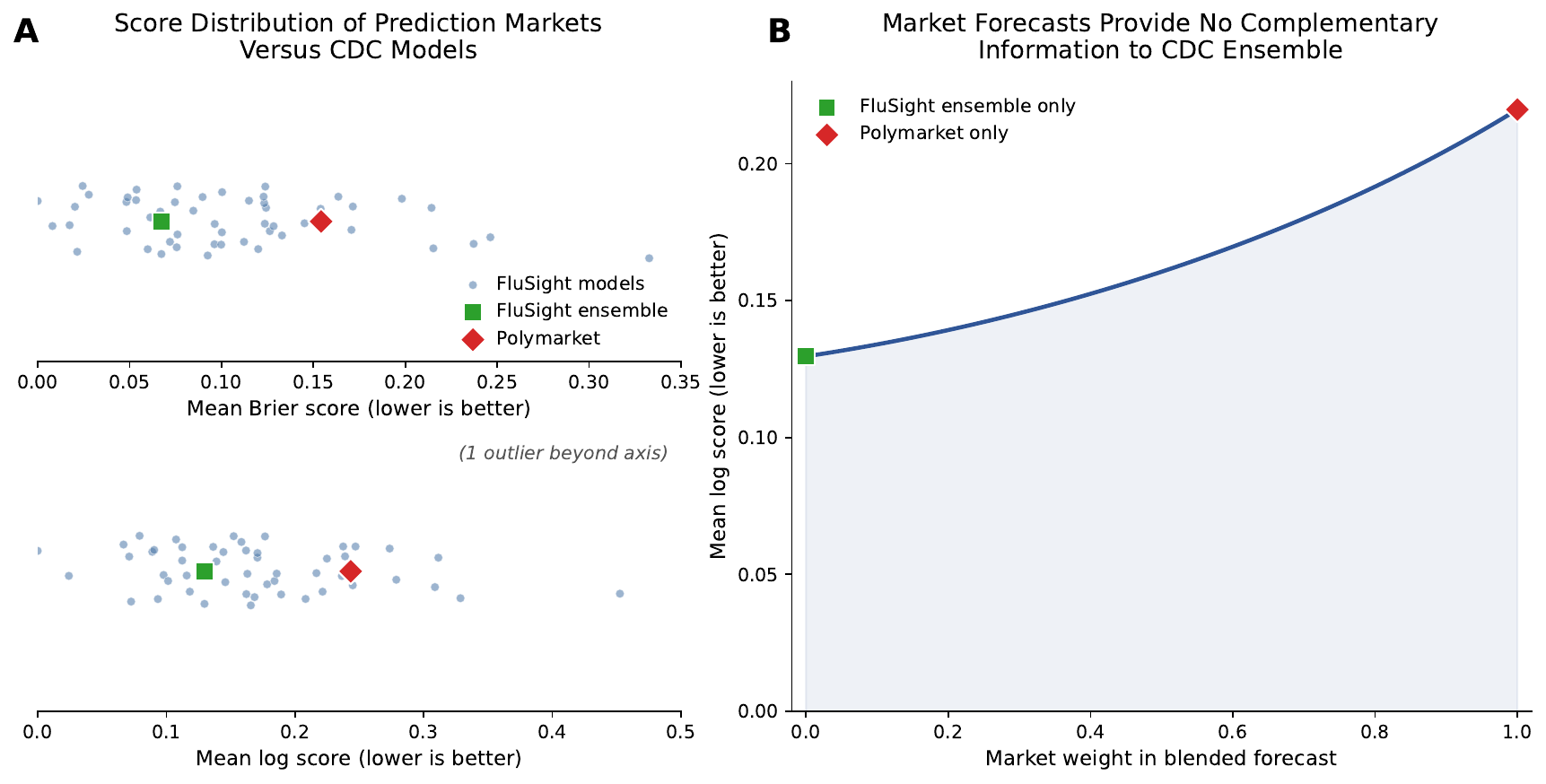}
\caption{Prediction market performance for weekly cumulative U.S. influenza hospitalizations, 2025--2026 season. (A) Distribution of Brier scores (left) and log scores (right) across FluSight Forecast Hub submissions. The prediction market (red) ranks at the 14th percentile by Brier score and the 23rd percentile by log score, placing it in the lower tail of participating models. The FluSight ensemble (blue) is shown for reference. (B) Score as a function of the ensemble weight $\alpha$ in the convex combination $q = \alpha \cdot p_{\text{flusight}} + (1 - \alpha) \cdot p_{\text{market}}$, evaluated under Brier score, log score, and CRPS. For Brier and log score, the optimal weight is $\alpha^* = 1.0$, indicating that the combined forecast places zero weight on the prediction market. For CRPS, the optimum at $\alpha^* = 0.82$ yields a negligible improvement over the ensemble alone.}
\label{fig:flu}
\end{figure}

Across 16 weekly cumulative influenza hospitalization markets resolving between January and May 2026, prediction markets achieved a mean Brier score of 0.154 and mean log score of 0.243. Relative to the distribution of FluSight Forecast Hub submissions, the prediction market ranked at the 17th percentile by both Brier score and log score (Figure~\ref{fig:flu}A), placing it in the lower tail of participating models. The FluSight ensemble outperformed the prediction market on every scoring rule considered.

A low percentile ranking does not rule out the possibility that prediction markets contribute information that is complementary in a way that is useful to the ensemble. Even a noisy forecasting model that achieves relatively higher error can be useful if it captures signal that other models miss (so that a weighted combination outperforms either forecast alone). We test whether prediction markets contribute complementary information by constructing a convex combination $q = \alpha \cdot p_{\text{flusight}} + (1 - \alpha) \cdot p_{\text{market}}$ and searching for the weight $\alpha^*$ that minimizes each scoring rule on the full sample of events. If the market carries information not in the ensemble, $\alpha^*$ will be less than one and the combined score will improve on the ensemble alone.

For both Brier and log score, $\alpha^* = 1.0$: the optimal combined forecast places zero weight on the prediction market and recovers the FluSight ensemble exactly (Figure~\ref{fig:flu}B). For CRPS, $\alpha^* = 0.82$ gave a mean score of 1.873 compared to 1.885 for the ensemble alone, a difference likely within expected sampling variability at this sample size. Prediction markets do not seem to contribute marginal predictive information conditional on the FluSight ensemble, meaning they should only be used cautiously as a complementary signal in forecasting systems.

\subsection{Measles markets are outperformed by a simple ARIMA baseline}

For measles there is no established forecasting hub, so we compare prediction markets against baseline models commonly used for forecasting. Across monthly cumulative measles case count markets spanning January through April 2026, prediction markets were outperformed by an auto-ARIMA baseline on every scoring rule and in every month of the study period. The ARIMA baseline achieved mean log scores roughly 34.4\% better than the prediction market, and the ordering held across all evaluated months. This is a setting in which prediction markets should have a clear advantage: the ARIMA baseline uses only past case counts, while market participants can draw on the same counts plus any additional source of information they choose, including news reports, local observations, and the output of more sophisticated models. That a univariate statistical model fit to public surveillance data outperforms the market is direct evidence that the market is not aggregating available information efficiently.

Taken together with the influenza results, prediction markets fail to achieve the performance that would justify their use as forecasting tools for infectious disease dynamics. For influenza, they are outperformed by the FluSight ensemble and do not contribute marginal information conditional on it; for measles, they are outperformed by univariate statistical models that can be fit in minutes.

\section{Discussion}

Prediction markets have a strong track record in other forecasting domains. In elections, economic indicators, and sports outcomes, they have repeatedly rivaled or outperformed expert forecasters and polls \cite{snowberg2012prediction, berg2008prediction}, and it makes sense that they would work: participants are financially incentivized to form accurate beliefs, can incorporate any information they deem relevant, and can update their forecasts in real time as new signals emerge \cite{berg2003accuracy}. Given this, it would be reasonable to expect markets to contribute useful forecasts for infectious disease dynamics. Our results suggest that, at present, they do not. For influenza, Polymarket forecasts sit in the lower tail of the FluSight Forecast Hub distribution and add no marginal information to the FluSight ensemble. For measles, markets are outperformed by an auto-ARIMA baseline fit to publicly available surveillance data that only uses historical counts. These findings hold despite an evaluation format that, if anything, advantages the markets.

One concrete manifestation of market inefficiency is the placement of probability mass on outcomes that are already ruled out by observed surveillance data. In cumulative forecasts, the target quantity is non-decreasing, so any probability assigned to values below the most recent observed cumulative count is guaranteed to be wrong. We observe this in practice: in the first flu market we analyze (resolving in the second week of 2026), 7.84\% of market-implied probability mass is placed on impossible bins. This fraction declines over the season as markets mature---2.46\% in week 3, 0.84\% in week 4, 1.04\% in week 5---but never reaches zero. An efficient market with sufficient liquidity and informed participants should arbitrage this mass to zero immediately. Its persistence is direct evidence that disease markets are not reaching the pricing efficiency observed in more established markets.

The failure is consistent with several differences from domains where markets have historically succeeded. Election and economic indicator markets attract sustained participation from informed traders with years of domain context and resolve on long horizons with relatively stable information flow, allowing prices to converge through repeated trading. Disease markets are new, have relatively less trading activity (the average week we analyze has $\$15,000-\$25,000$ in trading volume versus $\sim\$3.7$ billion for the 2024 election), attract participants whose epidemiological expertise is limited relative to the research groups contributing to forecasting hubs, and resolve on short horizons with rapidly changing ground truth. Disease forecasts also target continuous quantitative outcomes rather than the categorical outcomes markets have historically handled well. Even if these conditions could be addressed, doing so would require public health stakeholders to actively develop betting-based forecasting infrastructure, which raises ethical considerations that potentially discourage such efforts.

The development of betting-based forecasting infrastructure would have consequences beyond forecasting accuracy. From an incentives standpoint, more liquid disease markets would place more capital on the same outcomes that public health works to reduce. The concern also extends into expertise. Recruiting the epidemiologically informed traders that markets would require to perform well means drawing the people best positioned to improve surveillance toward betting infrastructure instead. Incorporating prediction markets into public health forecasting would also give commercial betting platforms a formal role in disease surveillance that would be difficult to undo. While these costs may not be prohibitive in principle, they're difficult to justify when forecasting systems with stronger empirical track records already exist.

The development of betting-based forecasting infrastructure would have consequences beyond forecasting accuracy. The performance gaps we document are at least partly attributable to low volume and limited participation by informed traders, and closing them would require deliberately growing disease-focused gambling markets and recruiting epidemiologically expert participants. This is uncomfortable on several fronts. Gambling itself is increasingly recognized as a public health concern \cite{Packin2026}, and it is hard to reconcile that view with public health institutions actively cultivating it. Recruiting the experts best positioned to improve surveillance into betting markets also has obvious opportunity costs. And there is a more basic discomfort with building forecasting infrastructure around wagers on human suffering. None of these costs are prohibitive in principle, but they are difficult to justify when forecasting systems with stronger empirical track records already exist.

Several limitations to this work apply: our evaluation covers only one platform, two diseases, and a single season, with small sample sizes (one year for measles and about half a year for flu). We focus on Polymarket because alternative platforms currently host few or no comparable infectious disease markets. As (if) other platforms introduce such contracts, and as existing markets accumulate trading history, later evaluations may yield different results. Despite these limitations, our findings are clear: current prediction markets underperform both ensembles and simple statistical baselines, exhibit measurable inefficiencies, and do not contribute complementary information to existing forecasting systems. Taken together, these results suggest that prediction markets should not be incorporated as components of infectious disease forecasting pipelines in their present form.

\bibliographystyle{unsrt}
\bibliography{references}

\end{document}